\documentclass[11pt]{article}

\usepackage[margin=2.5cm]{geometry}
\usepackage{amsthm}
\usepackage{amsmath}
\usepackage{amssymb}
\usepackage{amsfonts}
\usepackage{graphicx}
\usepackage{enumerate}
\usepackage{rotating}

\theoremstyle{definition}

\theoremstyle{remark}

\theoremstyle{plain}

\newcommand{\calo}{\mathcal{O}}

\newcommand{\N}{\textbf{N}}

\newcommand{\vp}{\vec{p}}
\newcommand{\vq}{\vec{q}}

\begin{document}
\title{On the AdS/QCD estimates of the scalar glueball mass}

\author{H.M. Ratsimbarison, \\
Institute of Astrophysics and High Energy Physics of Madagascar,\\
  Q 208, Faculty of Science Building,\\
  Antananarivo University}

\date{October 2009}
\maketitle

\begin{abstract}
We review standard holographic derivations of the scalar glueball spectra and emphasize 
computational assumptions which rely strongly on the original AdS/CFT correspondence, and those which are more 
arbitrary. In the soft-wall model, we show explicitly that the dual of the spectral mass constraint in the K\"allen-Lehmann representation of the 2-point scalar gluonic correlator is an eigenvalue constraint on the motion equation of the bulk field corresponding this correlator.
\end{abstract}

\section{Introduction}
In this paper, we review some extensions of the Anti de Sitter/Conformal Field Theory (AdS/CFT) correspondence to describe the gluonic sector of Quantum Chromodynamics (QCD). More precisely, we emphasize main concepts used in the holographic computation of the scalar glueball mass. By the way, the glueball mass analysis allows us to exhibit some advantages and limits of the AdS/QCD approach.

\section{General description of the AdS/CFT correspondence}
It is a conjectured correspondence between some conformal field theories on the conformal boundary M$_d$ of an Anti de Sitter space AdS$_{d+1}$ and string theories on the product of AdS$_{d+1}$ with a compact manifold. The correspondence allows to compute correlation functions of the boundary theory, defined on M$_d$, by means of appropriate partition functions of the bulk theory, defined on AdS$_{d+1}$. More explicitly, in the \emph{scalar sector} of the string theory, we have \cite{oaha99, juma97,edwi98}:
\begin{eqnarray}
 \left\langle0| T\left(e^{\int \phi_0\calo}\right)|0\right\rangle_{CFT} &=& \int_{\phi|_{M_d} = \phi_0}D\phi\; e^{iS_{string}[\phi]}, \label{adscftcor}
\end{eqnarray}
where the left hand side of (\ref{adscftcor}) is the generating functional of the conformal field theory, the rigth hand side is the supergravity (or string) partition function computed from bulk fields $\phi$ whose restriction at the boundary is $\phi_0$. In the classical approximation, we know that the partition function reduces to the exponential of the critical value of the classical action. Taking into account the boundary condition, the classical term of Z$_S$($\phi_0$) is given by the exponential of the critical value of the classical supergravity action I$_S$, evaluated with critical bulk field $\phi_{crit}$ with boundary value $\phi_0$,
\begin{align}
 \int_{\phi|_{M_d} = \phi_0}D\phi\; e^{iS_{string}[\phi]} = e^{iS_{SUGRA}[\phi_{crit}]} + \text{quantum and string corrections}.
\end{align}
Due to the lack of valuable computational tools for nonperturbative QCD, it will be worth to extend the AdS/CFT correspondence for QCD. Let us focus on the glueball spectrum analysis from the holographic point of view.
\section{About the mass of scalar glueball}
Before the holographic approach, let us recall that scalar glueball is bound state of gluons, with quantum numbers J$^{PC}$ = 0$^{++}$ predicted by QCD, but not yet observed experimentally, although some candidates exist \cite{stna97,vccm09}. In QCD, the 0$^{++}$ glueball mass is extracted from the gluonic 2-point correlator
\begin{eqnarray}
	\left\langle 0|T\left[J_s(x)J_s(0)\right]|0\right\rangle, \quad J_s := \beta(\alpha_s)Tr(G^2).
\end{eqnarray}
and its spectral representation, 
\begin{eqnarray}
 \Pi(q^2) &:=& i\int d^4x\;e^{iqx}\left\langle 0|T\left(J_s(x)J_s(0)\right)|0\right\rangle,\\
 &=& \displaystyle\sum_n \int \frac{d\vp}{2E_{\vp}} |\left\langle \Omega|J_s(0)|m_n,p_E\right\rangle|^2\left[\frac{\delta(\vq - \vp)}{q^0 - E_{\vp} + i\epsilon} + \frac{\delta(\vq + \vp)}{q^0 + E_{\vp} + i\epsilon}\right] = \int \frac{dt}{2\pi} \frac{\rho(t)}{q^2 - t + i \epsilon},
 \label{kalerep}
\end{eqnarray}
where $|$m$_n$,p$_E$) is glueball state with on-shell momentum p$_E$ = (E$_{\vp}$,$\vp$) and mass m$_n$, and $\rho$ the associated spectral function of glueball resonances. 
Extending the AdS/CFT correspondence to QCD would allow to compute QCD correlators from an analogous ansatz to (\ref{adscftcor}) with appropriate theory on a well-chosen bulk theory. The next section considers explicit holographic models of QCD, called \emph{AdS/QCD models}, in which the scalar glueball mass spectrum arises from Lorentz invariance constraint, which will be shown to be dual to the spectral constraint on the variable q in the K\"allen-Lehmann representation (\ref{kalerep}) of the gluonic correlator.

\section{Holographic approach to the scalar glueball spectrum}
Without proof \cite{edwi98}, let us assume that the 2-point scalar gluonic correlator is described by a massless scalar bulk field of Lagrangian
	\[S[\phi] = \int dvol_{bulk}\;g^{MN}\partial_{M}\phi\partial_{N}\phi,
\]
defined on an appropriate bulk space. In order to derive holographically the scalar glueball mass, our task is to fix the bulk space which can provide, by the intermediate of the bulk (critical) action, the generating functional of the gluonic correlator. In the so-called bottom-up approach, one fixes the bulk space geometry in such a manner that it can accommodate phenomenological results \cite{karetal06}: in general, the bulk space is a modified five dimensional Anti de Sitter space with metric, in Poincar\'e coordinates, such as  
\begin{eqnarray}
 && ds^2 = h(z)\frac{R^2}{z^2}\left(\eta_{\mu\nu}dx^{\mu}dx^{\nu} + dz^2\right) =: g_{MN}dx^Mdx^N \\
 \text{with}	&& ds^2 \sim ds^2_{AdS_5} = ds^2_{h(z) = 1} \quad \text{ for z }\rightarrow 0, \label{metcond}
\end{eqnarray}
where $\eta_{\mu\nu}$ is the Minkowski metric with signature (-,+++). Such space possesses conformal boundary localized at z $\rightarrow$ 0, where lives the boundary theory. In this settings, the Euler-Lagrange equation of the bulk field is obtained from the variational principle:
\begin{eqnarray}
	\delta S &=& \frac{1}{2}\int \sqrt{g}\left[g^{MN}\partial_M\delta\phi \partial_N\phi + g^{MN}\partial_M\phi \partial_N\delta\phi \right],\\
	&=& \int \sqrt{g}\left[g^{MN}\partial_M\delta\phi \partial_N\phi\right], \quad \text{(symmetricity of g)} \\
	&=& \int \left[\delta_M(\sqrt{g}g^{MN}\partial_{N}\phi\delta\phi) - \delta_M(\sqrt{g}g^{MN}\partial_N\phi)\delta\phi\right].
\end{eqnarray}
So
\begin{eqnarray}
  &&\left\{ 
\begin{aligned}
  & \delta S = 0\\
  & \delta \phi(x\in \text{Boundary space}) = 0 \\
  & \delta \phi \text{ arbitrary}
\end{aligned}
  \right. \quad \Rightarrow \quad \partial_M(\sqrt{g}g^{MN}\partial_N\phi) = 0,\\
  \Rightarrow && \partial_M(\sqrt{g}g^{MN})\partial_N\phi + \sqrt{g}g^{MN}\partial_M\partial_N\phi = 0,\\
  \Rightarrow && \partial_M\left(\frac{R^3h^{3/2}}{z^3}\right)\eta^{MN}\partial_N\int \frac{d^4x}{(2\pi)^4}e^{-iqx}\tilde{\phi}(q,z) + \frac{R^3h^{3/2}}{z^3}\eta^{MN}\partial_M\partial_N\int \frac{d^4x}{(2\pi)^4}e^{-iqx}\tilde{\phi}(q,z) = 0,\\
  && \text{where } \left(\frac{R^2h}{z^2}\right)^{-1}\eta^{MN} := g^{MN} \text{ and } g_{MN}g^{MN} = dim(bulk),\\
  \Rightarrow && \left(R^3\frac{3h^{1/2}\partial_zh}{2z^3} - \frac{3R^3h^{3/2}}{z^4}\right)\int \frac{d^4x}{(2\pi)^4}e^{-iqx}\partial_z\tilde{\phi} + R^3\frac{h^{3/2}}{z^3}\int \frac{d^4x}{(2\pi)^4}e^{-iqx}\left[-q^2\tilde{\phi} + \partial_z^2\tilde{\phi}\right] = 0, \\
  \Rightarrow && \int \frac{d^4x}{(2\pi)^4}e^{-iqx}\left[h(-q^2\tilde{\phi} + \partial_z^2\tilde{\phi}) + 3(\frac{\partial_zh}{2} - \frac{h}{z})\partial_z\tilde{\phi}\right] = 0, \\
  \Rightarrow && \partial_z^2\tilde{\phi} + 3\left(\frac{\partial_zh}{2h} - \frac{1}{z}\right)\partial_z\tilde{\phi} - q^2\tilde{\phi} = 0.
\end{eqnarray}
where 
	\[\tilde{\phi}(q,z) := \int d^4x\;e^{iqx}\phi(x,z).
\]
Particular choices of the metric function h satisfying the physical condition (\ref{metcond}) lead to specific models such as the soft-wall model \cite{karetal06,pcol07} where 
\begin{eqnarray}
 h(z) = e^{-\frac{2}{3}(cz)^2} , 
\end{eqnarray}
with c an inverse length which is a physical parameter of the model. In fact, the soft-wall metric function is chosen to make the bulk equation of motion into the following generalized Laguerre differential equation
\begin{eqnarray}
	&& \left[\partial^2_z - 3\left(\frac{2}{3}c^2z + \frac{1}{z}\right)\partial_z - q^2\right]\tilde{\phi}(q,z) = 0, 		\label{buleq} \\
	&\Leftrightarrow& \left\{ 
\begin{aligned}
  & [\partial^2_z + 2(\frac{\alpha + 1/2}{z} - c^2z)\partial_z - q^2 - (2\alpha +4)c^2]f(z) = 0, \\
  & f(z) := z^{-\alpha - 2}\tilde{\phi}(q,z) , \quad \alpha = 2.
\end{aligned}
  \right. 
\end{eqnarray}
According to proprieties of this type of differential equations, regular solutions $\tilde{\phi}_{crit}$ of (\ref{buleq}) have eigenvalues:
\begin{eqnarray}
	-q^2 = 4c^2(n + 2) \quad \text{with } n\in \N_+.
	\label{swmas}
\end{eqnarray}
Now, it suffices to show that one obtains exactly the same eigenvalue constraint on the variable of the Fourier gluonic correlator to argue that eigenvalues (\ref{swmas}) are identical to 0$^{++}$ glueball masses given by the K\"allen-Lehmann representation (\ref{kalerep}).\\
Indeed, let us write the critical bulk field in terms of its boundary restriction by means of a Green's function K, so
\begin{eqnarray*}
  && \phi_{crit}(x,z) =: \int d^4x'\;K(x-x',z)\phi_0(x') , \quad \phi_{crit}(x,z\rightarrow0) =: \phi_{0}(x), \\
  \text{and} && \tilde{\phi}_{crit}(q,z) = \int d^4x'\;e^{iqx'}\tilde{K}(q,z)\phi_0(x').
\end{eqnarray*}
where $\tilde{K}$ is necessarily a solution of the motion equation (\ref{buleq}) and satisfy the boundary condition 
\begin{align*}
	K(x-x',z\rightarrow 0) = \delta(x-x') \quad \text{or} \quad \tilde{K}(q,z\rightarrow0) = 1,
\end{align*}
which is equivalent to the fact that $\phi_{crit}(x,z\rightarrow0) = \phi_{0}(x)$.\\
Applying the AdS/CFT correspondence (\ref{adscftcor}), we have to evaluate the twice functional derivative of the critical action with rapport to the boundary field in order to obtain the gluonic correlator
\begin{eqnarray*}
  	&& \left\langle 0|T\left[J_s(x_1)J_s(x_2)\right]|0\right\rangle = \frac{\delta^2S[\phi_{crit}]}{\delta\phi_0(x_1)\delta\phi_0(x_2)},\\
  	&=& \frac{\delta^2}{\delta\phi_0(x_1)\delta\phi_0(x_2)}\int dvol_{bulk}\;g^{MN}\int dq\,\partial_M\left[e^{-iqx}\tilde{K}(q,z)\right]\int dx'\,e^{iqx'}\phi_0(x')\\
  	&& \hspace{2,9cm} \times \int dq'\,\partial_N\left[e^{-iq'x}\tilde{K}(q',z)\right]\int dx''\,e^{iq'x''}\phi_0(x''), \\
  	&=& \frac{\delta}{\delta\phi_0(x_2)}\int dvol_{bulk}\;g^{MN}\int dq\,\partial_M\left[e^{-iqx}\tilde{K}(q,z)\right]\int dx'\,e^{iqx'}\delta(x' - x_1)\\
  	&& \hspace{2,9cm} \times \int dq'\,\partial_N\left[e^{-iq'x}\tilde{K}(q',z)\right]\int dx''\,e^{iq'x''}\phi_0(x'') \\
  	&& +  \frac{\delta}{\delta\phi_0(x_2)}\int dvol_{bulk}\;g^{MN}\int dq\,\partial_M\left[e^{-iqx}\tilde{K}(q,z)\right]\int dx'\,e^{iqx'}\phi_0(x')\\
  	&& \hspace{2,9cm} \times \int dq'\,\partial_N\left[e^{-iq'x}\tilde{K}(q',z)\right]\int dx''\,e^{iq'x''}\delta(x'' - x_1), \\
  	&=& \int dvol_{bulk}\;g^{MN}\int dq\,e^{iqx_1}\partial_M\left[e^{-iqx}\tilde{K}(q,z)\right]\int dq'\,\partial_N\left[e^{-iq'x}\tilde{K}(q',z)\right]\int dx''\,e^{iq'x''}\delta(x'' - x_2) \\
  	&& +  \int dvol_{bulk}\;g^{MN}\int dq\,\partial_M\left[e^{-iqx}\tilde{K}(q,z)\right]\int dx'\,e^{iqx'}\delta(x' - x_2)\int dq'\,e^{iq'x_1}\partial_N\left[e^{-iq'x}\tilde{K}(q',z)\right], \\
  	&=& \int dvol_{bulk}\;g^{MN}\int dq\,e^{iqx_1}\partial_M\left[e^{-iqx}\tilde{K}(q,z)\right]\int dq'\,e^{iq'x_2}\partial_N\left[e^{-iq'x}\tilde{K}(q',z)\right] \\
  	&& + \int dvol_{bulk}\;g^{MN}\int dq\,e^{iqx_2}\partial_M\left[e^{-iqx}\tilde{K}(q,z)\right]\int dq'\,e^{iq'x_1}\partial_N\left[e^{-iq'x}\tilde{K}(q',z)\right], \\
  	&=& 2\int dvol_{bulk}\,\frac{R^3h^{3/2}}{z^3}[\int dq\,e^{iqx_1}\left[(-iq^{\mu})\tilde{K}(q,z)\right]\int dq'\,e^{iq'x_2}e^{-i(q+q')x}\left[(-iq'_{\mu})\tilde{K}(q',z)\right] \\
  	&& \hspace{3,3cm} + \int dq\,e^{iqx_1}\left[\partial^z\tilde{K}(q,z)\right]\int dq'\,e^{iq'(x_2-x)}\left[\partial_z\tilde{K}(q',z)\right]],
  	\end{eqnarray*}
  	\begin{eqnarray*}
  	&=& 2\int dvol_{bulk}\,\frac{R^3h^{3/2}}{z^3}[\int dq\,e^{iqx_1}\left[(-iq^{\mu})\tilde{K}(q,z)\right]\int dq'\,e^{iq'x_2}e^{-i(q+q')x}\left[(-iq'_{\mu})\tilde{K}(q',z)\right] \\
  	&& \hspace{3,3cm} + \int dq\,e^{iqx_1}\left[\partial_z\tilde{K}(q,z)\right]\int dq'\,e^{iq'x_2}e^{-i(q+q')x}\left[\partial_z\tilde{K}(q',z)\right]],\\
  	&=& 2\int dz\,\frac{R^3h^{3/2}}{z^3}[\int dq\,e^{iqx_1}\left[(-iq^{\mu})\tilde{K}(q,z)\right]\int dq'\,e^{iq'x_2}\delta(q+q')\left[(-iq'_{\mu})\tilde{K}(q',z)\right] \\
  	&& \hspace{3,3cm} + \int dq\,e^{iqx_1}\left[\partial_z\tilde{K}(q,z)\right]\int dq'\,e^{iq'x_2}\delta(q+q')\left[\partial_z\tilde{K}(q',z)\right]],\\
  	&=& 2\int dz\,\frac{R^3h^{3/2}}{z^3}\left[\int dq\,e^{iq(x_1 - x_2)}q^2\tilde{K}^2(q,z) + \int dq\,e^{iq(x_1 - x_2)}\left(\partial_z\tilde{K}(q,z)\right)^2\right],\\
  	&=& \int dq\,e^{iq(x_1 - x_2)}\int dz\,\frac{2R^3h^{3/2}}{z^3}\left[q^2\tilde{K}^2(q,z) + \left(\partial_z\tilde{K}(q,z)\right)^2\right],\\
  	&=& \int dq\,e^{iq(x_1 - x_2)}\int dz\,\frac{2R^3h^{3/2}}{z^3}\left[q^2\tilde{K}^2(q,z) + \partial_z\left(\tilde{K}\partial_z\tilde{K}\right)(q,z) - \tilde{K}\partial_z^2\tilde{K}(q,z)\right].
\end{eqnarray*}
Therefore, the Fourier transform of the 2-point gluonic correlator is given by:
\begin{eqnarray}
	\Pi(q^2) &:=& i\int d^4x\;e^{iqx}\left\langle 0|T\left(J_s(x)J_s(0)\right)|0\right\rangle,\\
	&=& i\int dz\,\frac{2R^3h^{3/2}}{z^3}\left[q^2\tilde{K}^2(q,z) + \partial_z\left(\tilde{K}\partial_z\tilde{K}\right)(q,z) - \tilde{K}\partial_z^2\tilde{K}(q,z)\right].
\end{eqnarray}
One can go further by using the fact that $\tilde{K}$ is a solution of (\ref{buleq}) and the derivation result
\begin{eqnarray}
	\partial_z\left(\frac{R^3h^{3/2}}{z^3}\right) = 3\frac{R^3h^{3/2}}{z^3}\left(\frac{\partial_zh}{2h} - \frac{1}{z}\right).
\end{eqnarray}
So 
\begin{eqnarray*}
	 \Pi(q^2) &=& i\int dz\,\left\{\frac{2R^3h^{3/2}}{z^3}\left[q^2\tilde{K}^2 - \tilde{K}\partial_z^2\tilde{K}\right]  + \partial_z\left(\frac{2R^3h^{3/2}}{z^3}\tilde{K}\partial_z\tilde{K}\right) - \partial_z\left(\frac{2R^3h^{3/2}}{z^3}\right)\tilde{K}\partial_z\tilde{K}\right\},\\
	 &=& i\int dz\,\left\{\frac{2R^3h^{3/2}}{z^3}\tilde{K}\left[q^2\tilde{K} - \partial_z^2\tilde{K} - 3\left(\frac{\partial_zh}{2h} - \frac{1}{z}\right)\partial_z\tilde{K}\right] + \partial_z\left(\frac{2R^3h^{3/2}}{z^3}\tilde{K}\partial_z\tilde{K}\right) \right\},\\
	&=& i\int dz\, \partial_z\left(\frac{2R^3h^{3/2}}{z^3}\tilde{K}\partial_z\tilde{K}\right)(q,z),\\
	&=& \left[i\frac{2R^3h^{3/2}}{z^3}\tilde{K}\partial_z\tilde{K}(q,z)\right]^{z\rightarrow +\infty}_{z\rightarrow 0},\quad \text{with } -q^2 = 4c^2(n + 2) ,\quad n\in \N_+,
\end{eqnarray*}
where we obtain the holographic constraint on the variable of the Fourier transform $\Pi$ in the last equality.
The obtained 0$^{++}$ mass spectrum (\ref{swmas}) is comparable to linear dependence of rho meson masses with rapport to high radial quantum number \cite{karetal06}. Here are more general remarks deduced partially from the above analysis:
\begin{itemize}
	\item in the limit c $\rightarrow$ 0 or equivalently for pure Anti de Sitter bulk space, the boundary theory is massless, and therefore conformally invariant as confirmed by the AdS/CFT correspondence: we can conclude that the bulk metric encodes the conformality or not of the boundary theory;
	\item although phenomenological arguments allow to construct QCD dual in AdS/QCD models, the so-called top-down approach, which start directly from the full machinery of string theories, provides more controllability on the construction of QCD dual.
\end{itemize}

\section{Conclusion}
This review emphasizes some details on the identification of glueball mass spectrum provided by holographic models, such as the dual of the spectral constraint given by the K\"allen-Lehmann representation. However, more works need to be done to integrate further physical proprieties such as confinement and mass gap in AdS/QCD models. Our future work on AdS/QCD models is to fix the dual of supersymmetry in the Maldacena duality in order to suppress its effects on the boundary theory, to classify eigenvalue constraints on q$^2$ with rapport to the metric function, and finally to compare the holographic gluonic 2-point correlator with QCD results in order to impose more criteria on the choice of the metric function.

\paragraph{Acknowledgments.} I am grateful to Pr Stephan Narison for introducing me to this topic. Comments and discussions with Fenompanirina Andrianala was also benefit to the present work.


\begin{thebibliography}{20}
\bibitem{stnabk02} S. Narison, \emph{QCD as a theory of hadrons: from partons to quarks}, Cambdrige Press Edition, 812p, \textbf{2002}.
\bibitem{stna97} S. Narison, \emph{Masses, decays and mixings of gluonia in QCD}, ArXiv preprint: hep-ph/9612457, \textbf{1997}.
\bibitem{vccm09} V. Crede, C.A. Meyer, \emph{The experimental status of glueballs}, Arxiv preprint: 0812.0600 [hep-ex], \textbf{2009}.
\bibitem{gafe08} F. Ferretti, \emph{A very basic introduction to the AdS/CFT correspondence}, ICTP Introductory School on Gauge Theory/Gravity Correspondence, \textbf{2008}.
\bibitem{juma97} J. Maldacena, \emph{The Large N limit of superconformal field theories and supergravity}, \textit{Adv. Theor. Math. Phys.} \textbf{2} (1998) 231, hep-th/9711200.
\bibitem{edwi98} E. Witten, \emph{Anti-de Sitter space and holography}, \textit{Adv. Theor. Math. Phys.} \textbf{2} (1998) 253, arXiv preprint: hep-th/9802150.
\bibitem{oaha99} O. Aharony, S. Gubser, J. Maldacena, H. Ooguri, and Y. Oz, \emph{Large N field theories, String theory and Gravity}, \textit{Phys. Rept.} \textbf{323}, 183-386 (2000), hep-th/9905111.
\bibitem{ccmr06} C. Csaki, M. Reece, \emph{Toward a systematic holographic QCD; a braneless approach}, arXiv preprint: hep-ph/0608266, \textbf{2006}.
\bibitem{karetal06} A. Karch, E. Katz, D. T. Son, M. A. Stephanov, \emph{Linear confinement and AdS/QCD}, arXiv preprint: hep-ph/0602229, \textbf{2006}.
\bibitem{pcol07} P. Colangelo, F. De Fazio, F. Jugeau, S. Nicotri \emph{On the light glueball spectrum in a holographic description of QCD}, arXiV preprint: hep-ph/0703316, \textbf{2007}.
\bibitem{hfor08} H. Forkel, \emph{Glueball correlators as holograms}, arXiv preprint: 0808.0304 [hep-ph], \textbf{2008}.
\bibitem{snic08} S. Nicotri, \emph{Scalar glueball in a holographic model of QCD}, arXiv preprint: 0807.4377 [hep-ph], \textbf{2008}.
\end{thebibliography}
\end{document}